\documentclass[
    reprint,         % Two-column format (like PRL)
    aps,             % American Physical Society style
    prl,             % Journal: Physical Review Letters
    superscriptaddress, % Author affiliations in superscripts
    showpacs,        % Show PACS numbers (optional)
    floatfix,        % Fixes for floating figures/tables
    nofootinbib      % Keeps footnotes out of bibliography
]{revtex4-2}

\usepackage[utf8]{inputenc}    % UTF-8 encoding
\usepackage[T1]{fontenc}       % Better font encoding
\usepackage{amsmath, amssymb}  % Math symbols
\usepackage{bm}                % Bold math
\usepackage{graphicx}          % Include graphics
\usepackage{xcolor}            % Color text
\usepackage{hyperref}          % Hyperlinks
\usepackage{physics}           % Common physics notations (\bra, \ket, etc.)
\usepackage{siunitx}           % SI units formatting
\usepackage{microtype}         % Better text spacing
\usepackage{newtxtext}         % PRL-like font
\usepackage{newtxmath}         % PRL-like font
\usepackage{multirow}          % For tables
\usepackage{threeparttable}    % For tables
\usepackage[version=3]{mhchem} % Formula subscripts using \ce{}
\usepackage{soul}

\hypersetup{
    colorlinks=true,
    linkcolor=blue,
    citecolor=blue,
    urlcolor=blue,
    pdfauthor={Your Name},
    pdftitle={Your Title},
}

\begin{document}

\title{Chemistry in a Cryogenic Buffer Gas Cell}

\author{Qi Sun}
\altaffiliation{These authors contributed equally to this work}
\affiliation{Department of Physics, Columbia University, New York, New York 10027, USA}

\author{Jinyu Dai}
\altaffiliation{These authors contributed equally to this work}
\altaffiliation{\\{\hypersetup{urlcolor=blue}\href{mailto:jd3706@columbia.edu}{jd3706@columbia.edu}}}
\affiliation{Department of Physics, Columbia University, New York, New York 10027, USA}

\author{Rian Koots}
\altaffiliation{These authors contributed equally to this work}
\affiliation{Department of Physics and Astronomy, Stony Brook University, Stony Brook, New York, New York 11794, USA}

\author{Benjamin C. Riley}
\affiliation{Department of Physics, Columbia University, New York, New York 10027, USA}

\author{Jes\'us P\'erez-R\'ios}
\email{jesus.perezrios@stonybrook.edu}
\affiliation{Department of Physics and Astronomy, Stony Brook University, Stony Brook, New York, New York 11794, USA}

\author{Debayan Mitra}
\email{debayanm@iu.edu}
\affiliation{Department of Physics, Columbia University, New York, New York 10027, USA}
\affiliation{Department of Physics, Indiana University, Bloomington, Indiana 47405, USA}

\author{Tanya Zelevinsky}
\email{tanya.zelevinsky@columbia.edu}
\affiliation{Department of Physics, Columbia University, New York, New York 10027, USA}

\date{\today}

\begin{abstract}
% \noindent
Cryogenic buffer gas sources are ubiquitous for producing cold, collimated molecular beams for quantum science, chemistry, and precision measurements.  The molecules are typically produced by laser ablating a metal target in the presence of a donor gas. The radical of interest emerges due to a barrier-free reaction or under thermal or optical excitation. High-barrier reactions, such as between Ca and H$_2$, should be precluded. We study chemical reactions between Ca and three hydrogen isotopologues H$_2$, D$_2$, and HD in a cryogenic cell with helium buffer gas. We observe that H$_2$ can serve as both a reactant and a buffer gas, outperforming D$_2$ and HD. We use a reaction network model to describe the chemical dynamics and find that the enhanced molecular yield can be attributed to rapid vibrational excitations of the reactant gas. Our results demonstrate a robust method for generating bright cold beams of alkaline-earth-metal hydrides for laser cooling and trapping.
\end{abstract}

\maketitle

% \section{Introduction}

Chemical reactions at low densities and low relative energies are characterized by the limited number of available reactive trajectories. Such reactions occur in astrophysical environments and may be responsible for the production of hydrocarbons in the early universe~\cite{DALGARNO19871,Smith_Astrochem_review}. Studies of rate coefficients for single-collision events in controlled laboratory environments can shed light on phenomena such as quantum interference between reaction pathways~\cite{Xie_Quantum_Interference, son_ketterle_2022_science_interference} and the role of entanglement in reactions~\cite{Liu_Ni_entanglement}. Techniques such as cold molecular beams~\cite{Lee_1987_Science_molecular_beam, Jankunas_review_cold_beam}, supersonic jets~\cite{Zare_supersonic, kilaj_willitsch_2018_natcomm_supersonicjet, vandeMeerakker2012}, Stark deceleration~\cite{bethlem1999_PRL_StarkDecelerator, Jankunas_review_cold_beam}, centrifuge deceleration~\cite{Rempe_cryofuge}, ion trapping~\cite{Merkt_ion_molecules,Nerevicius_feshbach_tomography, staanum2010_NatPhys_MoleculeIonTrap, xu_willitsch_2024_PRL_MoleculeIonTrap,Hirzler_2022}, optical lattice trapping~\cite{ZelevinskyMcDonaldNature16_Sr2PD}, and optical tweezers~\cite{Yu_2022_chem_review} have enabled extreme reaction regimes unrealizable with traditional techniques and have exposed new quantum phenomena~\cite{Bohn_ColdMol_2017}.

However, there is an intermediate regime in the chemical reaction landscape where densities and temperatures straddle the line between quantum and classical~\cite{ZelevinskyKondovPRL18_PDQuantumQuasiclassicalXover}. Here, the motional states of the reactants can be treated classically but the quantum nature of their internal states cannot be neglected. This regime is manifested in cryogenic buffer gas beam (CBGB) sources~\cite{DeLucia_BufferGas_1984,Doyle_CBGB_PRL_2005, Hutzler_CR2012_BufferGasBeams}. Typical gas densities in a CBGB source can be of the order of $10^{15}$~cm$^{-3}$ and equilibrium temperatures of the order of 4~K. But laser ablation, which is typically used to create the molecules of interest, can produce transient temperatures as high as $10^4$~K~\cite{Zelevinsky_BaH_Spectroscopy_PRA_2016}. Hence this configuration allows scattering processes that extend from multiple partial waves~\cite{Bohn_scattering_PRA_2000,Klos_C60_2024} to simple hard-sphere scattering~\cite{Weinstein_Li_CaH_reaction_PRL_2012}. CBGB sources are experimentally versatile since different gaseous reactants can be introduced with ease~\cite{Wright17092023}. Therefore, a CBGB source is an ideal playground to study chemistry over a large parameter range~\cite{Truppe_CaF_source_2018,Jadbabaie_YbOH_enhancement_2020,Albrecht_BaF_CBGB_2020,Mooij_BaF_beam_2024}. 

In a CBGB source, three physical and chemical processes can occur. The first process involves elastic collisions between particles of interest and the cold buffer gas, which result in translational cooling. The second process is inelastic collisions that lower the internal temperature, thereby achieving ro-vibrational cooling. The third process consists of chemical reactions responsible for forming the desired particles. The first two processes together are known as buffer gas cooling, a technique that has proven highly effective in generating cold atomic and molecular beams in their ro-vibrational ground states~\cite{Hutzler_CR2012_BufferGasBeams}. The third, reactive process has attracted renewed attention as many new molecular species have been successfully produced in CBGBs. These molecules range from diatomics such as CaF~\cite{Anderegg_RFMotCaF_2017}, YbF~\cite{Tarbutt_YbF_PRL_2018} and AlF~\cite{Wright17092023,Truppe_2019}, to triatomics such as CaOH~\cite{Vilas_2022_3D_MOT_CaOH}, YbOH~\cite{Augenbraun_NJP2020_YbOH_Sisyphus} and SrOH~\cite{Kozyryev_SisyphusSrOH_2017}, to polyatomics such as CaOCH$_3$~\cite{Mitra_CaOCH3Sisphus_2020} and CaOC$_6$H$_5$~\cite{Zhu_aromatic_NatChem_2022}. To produce a molecule of type $MX$, the common approach involves laser ablation of a metal target ($M$ = Ca, Sr, Yb, $\ldots$), followed by allowing the resulting hot plume to chemically react with a donor gas ($RX$ = SF$_6$, H$_2$O, CH$_3$OH, $\ldots$).

Reactions occurring within CBGB sources are generally barrierless or possess small energy barriers that can be overcome with thermal or optical excitation. For instance, Ca reacts exothermically with SF$_6$ to form CaF~\cite{10.1063/5.0098378}, whereas its reaction with H$_2$O to yield CaOH necessitates promoting the Ca atom to the metastable $^3P_1$ state in order to overcome an energy barrier of approximately $1.3$~eV~\cite{Vilas_2022_3D_MOT_CaOH}. By contrast, the Ca + H$_2$ reaction has a considerably higher barrier of $\sim$$2.7$~eV. To thermally overcome this barrier, temperatures on the order of $3\times10^4$~K would be needed, which is approaching the limits achievable through laser ablation. Consequently, one would anticipate the reaction to yield only a small amount of product. In this work, we investigate the reactions between Ca atoms and the isotopologues H$_2$, D$_2$, and HD, under conditions both including and excluding helium as the buffer gas. This reaction has been studied extensively at high temperatures and pressures~\cite{Albert_CaH_1909,Watson_CaH_CaD_1935,Steimle_CaH_2004}, conditions that preclude the cryogenic environment of a CBGB source. The formation of CaH (or CaD) is of particular interest for the production of ultracold hydrogen (or deuterium). The molecule CaH (or CaD) is more favorable for laser cooling compared to atomic H (or D)~\cite{Vazquez-Carson_2022,Dai2024PRR_CaD} and the cooled and trapped molecule can potentially be dissociated near the threshold~\cite{Sun_PRR2023_Predissociation}. Counterintuitively, our results show that the reaction proceeds much more efficiently than a simple thermodynamic estimate would suggest, emphasizing that reaction dynamics play a critical role compared to purely energetic considerations. Furthermore, we observe that H$_2$, but not D$_2$ and HD, acts effectively as a buffer gas, enabling the produced CaH molecules to thermalize into their ro-vibrational ground state. To interpret these observations, we developed a model of the reactant dynamics employing rate constants obtained from quasi-classical trajectory (QCT) simulations. Our calculations elucidate the general trends observed in the experiment and provide an order-of-magnitude estimate of reaction rate constants. Additionally, the model serves to predict loss effects in the buffer gas cell.

%\section{Experimental Implementation}

\begin{figure}[t]
   \centering
   \includegraphics[scale=0.42]{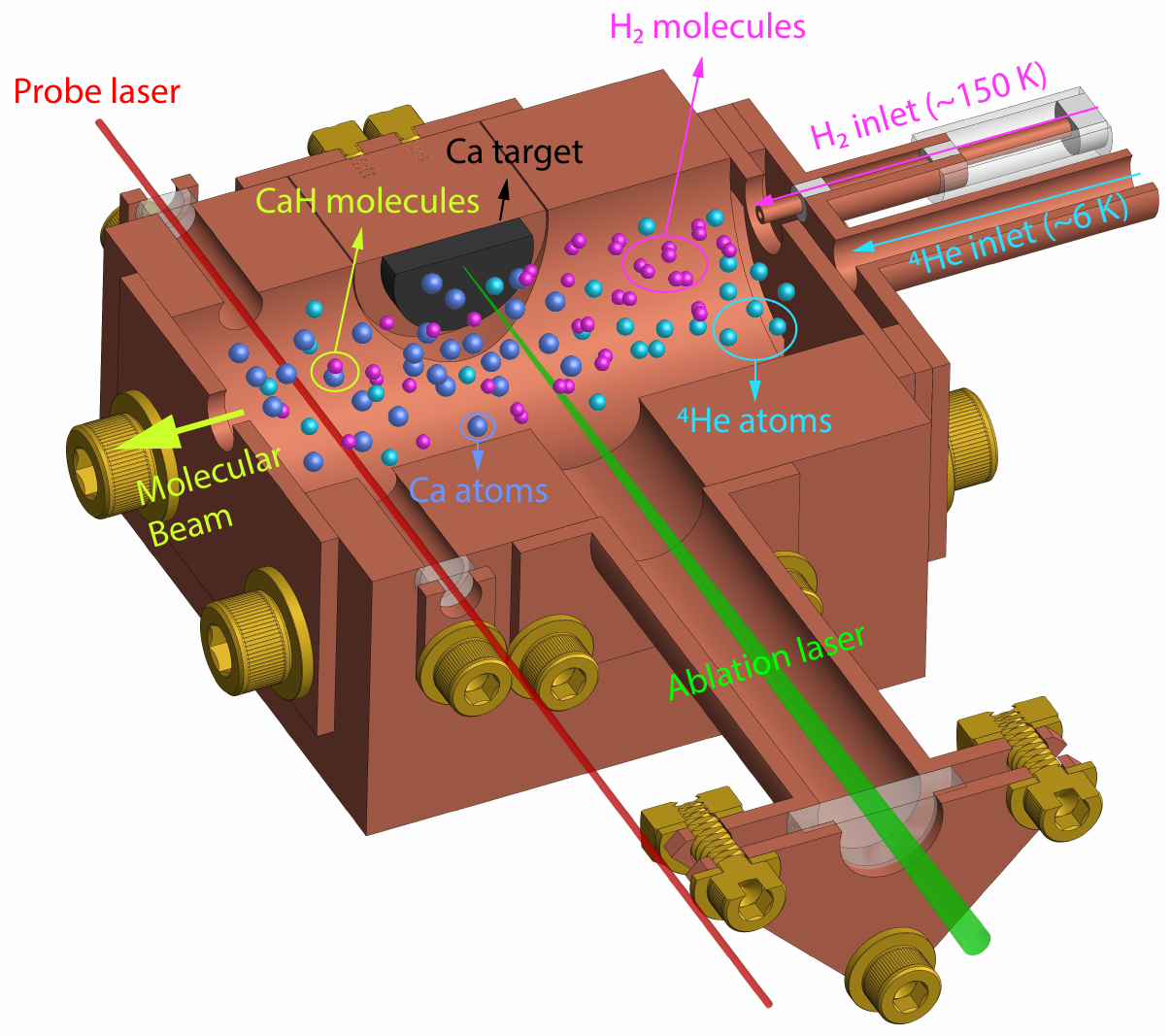}
   \caption{Horizontal cross section of the cryogenic buffer gas cell. Hot Ca atoms are generated via laser ablation of a solid target and subsequently react with H$_2$ molecules that flow in at $\sim$$150$~K. Additionally, $^4$He flows in at $\sim$$6$~K for efficient buffer gas cooling. The Ca atoms and the product molecules are probed through laser absorption, with optical access $\sim$$2$~cm downstream from the target.}
   \label{fig:setup}
\end{figure}

The experimental setup is illustrated in Fig.~\ref{fig:setup}, similar to that described in Ref.~\cite{Hutzler_CR2012_BufferGasBeams}. The apparatus consists of a copper cell with internal dimensions of $25.4$~mm in diameter and $50.8$~mm in length, thermally anchored to the cold plate of a pulse tube refrigerator maintained at $\sim$$6$~K. A calcium metal target is mounted inside the cell. A Q-switched Nd:YAG laser operating at $532$~nm is focused onto the calcium target with a beam diameter of roughly $100$~$\mu$m. Each ablation pulse carries up to $30$~mJ of energy, lasts about $10$~ns, and repeats at $\sim$$1$~Hz, creating a hot plume of calcium atoms. Multiple species of gas are flown simultaneously into the cell to facilitate chemical reactions and thermalization. A thermally isolated fill line, held at $\sim$$150$~K, delivers H$_2$, D$_2$, or HD gases into the cell. Helium is pre-cooled to $\sim$$6$~K and diffused to achieve a homogeneous distribution inside the cell for more effective cooling. Laser absorption spectroscopy is used to probe the target atoms and molecules through an optical window located $\sim$$2$~cm downstream from the ablation site. Specifically, calcium atoms are probed on the $4s4p$ $^3P_1\leftarrow4s^2$ $^1S_0$ transition at $657$~nm, and CaH molecules are detected via the $A^2\Pi_{1/2}(\nu'=0,\ J'=3/2,\ +)\leftarrow X^2\Sigma^+(\nu''=0,\ N''=1,\ J''=1/2,\ -)$ transition at $695$~nm. The measured CaH molecular density in the $X^2\Sigma^+(\nu=0,\ N=1,\ J=1/2,\ -)$ state is used to calculate populations across all ro-vibrational levels up to ($\nu=0,\ N=2$), thereby accounting for all the cold molecules produced (see Supporting Information).

%\section{Results}

%\subsection{Ca + H$_2$ chemical reaction in a CBGB source}

\begin{figure}[t]
   \centering
   \includegraphics[scale=0.61]{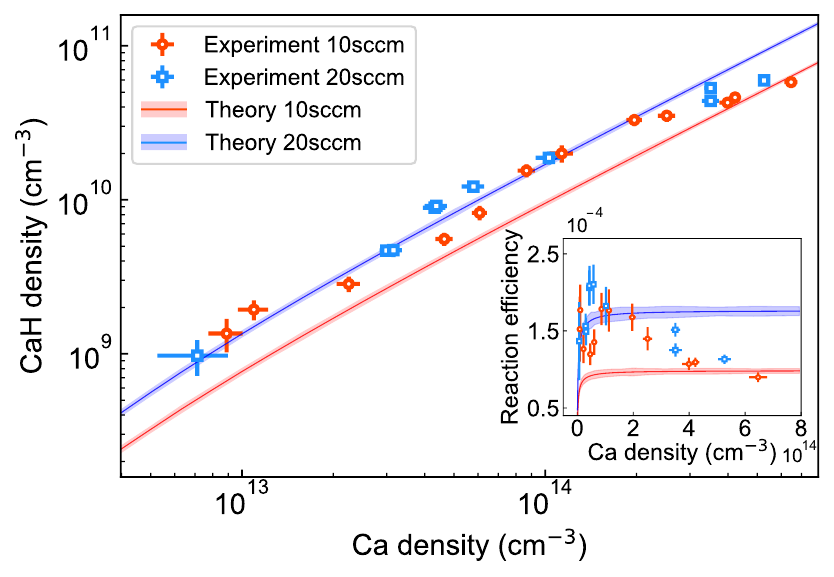}
   \caption{Chemical reaction in a cryogenic buffer gas cell. Time-averaged CaH densities measured within 0--1~ms after ablation, plotted against calcium density for $10$~SCCM of H$_2$ (red circles) and $20$~SCCM of H$_2$ (blue squares), both under $2.2$~SCCM of He. The lines denote theoretical results at the same conditions at a collision temperature of $2400$~K, showing qualitative agreement. The inset shows the reaction efficiency, defined as the ratio of CaH to calcium densities. A notable decrease in reaction efficiency is observed experimentally at higher calcium densities, whereas the theoretical model does not capture this trend, showing the need for further refinement. Error bars on the experimental data correspond to 1-$\sigma$ statistical uncertainties.}
   \label{fig:reaction}
\end{figure}

The Ca + H$_2$ chemical reaction is expected to be barely within reach in a CBGB source due to its endothermic nature. Surprisingly, we detect a substantial yield of CaH molecules ($\sim$$3$--$10 \times 10^{10}$ molecules$/$steradian$/$pulse), which we attribute to the unique properties of cryogenic H$_2$, as discussed below. We carefully calibrate the internal state distribution at different ablation energies up to the highest populated level ($\nu=0,\ N=2$) (see Supporting Information) for the experimental conditions in Fig.~\ref{fig:reaction}, thereby concluding that the total CaH population across these states accounts for the total molecular yield. The reaction occurs once the H$_2$ flow reaches a few standard cubic centimeters per minute ($1~\text{SCCM}\approx4.5\times10^{17}$ molecules per second) and plateaus at flow rates of several tens of SCCM. In comparison, this process is $\sim$2--3 orders of magnitude more efficient than the exothermic reactions involving the two other hydrogen isotopologues studied. While $2.2$~SCCM of helium is introduced in the experiments shown in Fig.~\ref{fig:reaction}, our results further demonstrate that H$_2$ alone can function as an effective buffer gas, eliminating the necessity of flowing helium.

Figure~\ref{fig:reaction} shows the measured total CaH density plotted against the initial calcium density. The calcium density is controlled by tuning the ablation energy, which simultaneously influences the temperature of the ablation plume. We develop a chemical reaction network model (lines in Fig.~\ref{fig:reaction}) based on rate constants obtained from QCT calculations (see Supporting Information). We find a constant plume temperature of $\sim$$2400$~K, independent of calcium density, yields the best fit to the data. We verify this assumption by measuring the plume temperature without gas flow (see Supporting Information). Our measurements show a maximum $45(10)$\% increase in temperature with ablation energy, supporting the constant plume temperature assumption. We also note that plume temperature is the only fit parameter we use in the reaction network model.

We observe that the experimental data at different H$_2$ flow rates in Fig.~\ref{fig:reaction} appear less well-separated than the theoretical predictions. In our model, we assume that the H$_2$ density increases linearly with flow rate, which is a simplification of the actual processes occurring in the cell. In reality, a significant portion of H$_2$ may condense onto the cold cell walls, forming solid H$_2$ ice, while sublimation from this ice layer can reintroduce H$_2$ into the gas phase. These dynamics may lead to transient and spatially inhomogeneous gas densities, but they are difficult to model and lie beyond the scope of this work. Experimentally, we also observe that CaH production remains substantial even after the H$_2$ flow is turned off, further supporting the presence of residual H$_2$ vapor from sublimation. These effects may contribute to the reduced sensitivity of CaH yield to the H$_2$ flow rate observed in the experiment.

Although the experimental data generally agree with the reaction model predictions, noticeable deviations arise at higher calcium densities. As shown in the inset of Fig.~\ref{fig:reaction}, the reaction efficiency, or the fraction of calcium atoms converted into CaH molecules, decreases as the calcium density increases. In contrast, the model predicts that the reaction efficiency should plateau at higher densities. These losses may be attributed to non-equilibrium processes that are not included in the simplified model, for instance, the formation of denser ionic species such as Ca$^+$ within the plasma can enhance the likelihood of subsequent ionization of CaH, thereby reducing the overall yield. We explore the effect of Ca$^+$ ion production in the model to characterize these losses (see Supporting Information). The model addresses reactive, inelastic, and elastic collisions in the buffer gas cell.

%\subsection{Reaction network model}
%\label{sec:network}

In the reaction model, hot neutral calcium atoms collide with H$_2$ molecules, leading to vibrational excitation of H$_2$ and formation of CaH. While excited calcium atoms are likely to be produced in the ablation plume, they have lifetimes on the order of $1$~$\mu$s due to collisions with surrounding neutral atoms, much shorter than the reaction times considered in this work. Although H$_2$ flows into the cell at $150$~K, it rapidly thermalizes via collisions with He to $6$~K where nearly all the molecules are in the ground rovibrational state. The cold H$_2$ gas in the CBGB source is assumed to be in the ground vibrational state initially. Assuming a homogenous mixture of gases, the vibrational quenching via inelastic collisions can be represented by
\begin{equation}\label{eq:excite}
    \ce{H2($\nu$) + Ca <=>[$k^q_{\nu\rightarrow \nu^{\prime}}$($T_{\ce{Ca}}$)][$k^q_{\nu^{\prime}\rightarrow \nu}$($T_{\ce{Ca}}$)] H2($\nu^{\prime}$) + Ca,} 
\end{equation}
where up to ($\nu^{\prime} = 2$) are included. The reactive process representing the formation of CaH in all rovibrational states is described as
\begin{equation}\label{eq:react}
    \ce{H2($\nu$) + Ca ->[$k^r_{\nu}$($T_{\ce{Ca}}$)] CaH + H,}
\end{equation}
where $\nu$ represents the vibrational level of H$_2$, with up to ($\nu = 2$) included. The state-to-state quenching rate constants ($k^{q}$) and reaction rate constants ($k^{r}$) are calculated via the QCT method (see Supporting Information). Rotational excitation is neglected in this work, as our simulation shows that vibrational excitation has a more significant effect on the reaction rate than rotational excitation. This can be attributed to the far more efficient vibration-translation coupling compared to rotation-translation coupling~\cite{10.1063/5.0241219}. The inelastic rate constants, summed over all final rotational states, and the reactive rate constants, summed over all final rovibrational states, are shown in Fig.~S3(a) of the Supporting Information.

We use a deterministic approach to study the evolution of these reactions in a closed system at thermal equilibrium by solving the set of differential equations associated with Eqs.~(\ref{eq:excite})--(\ref{eq:react}). The evolution of CaH and H$_2$ vibrational state densities is given by
\begin{equation}\label{eq:dcahdt}
    \frac{d n_{\ce{CaH}}}{d t} = \sum_{\nu=0}^{2}k^r_\nu n_{\ce{H2}(\nu)}n_{\ce{Ca}}
\end{equation}
and
\begin{equation}\label{eq:dh2dt}
\begin{aligned}
     \frac{d n_{\ce{H2}(\nu)}}{d t} = &- k^r_{\nu} n_{\ce{H2}(\nu)}n_{\ce{Ca}} +\\ &\sum_{\nu^{\prime} \neq \nu}(k^q_{\nu^{\prime}\rightarrow \nu}n_{\ce{H2}(\nu^{\prime})}n_{\ce{Ca}} - k^q_{\nu\rightarrow \nu^{\prime}}n_{\ce{H2}(\nu)}n_{\ce{Ca}}),
\end{aligned}
\end{equation}
where  $n_{\ce{H2}(\nu)}$, $n_{\ce{CaH}}$, and $n_{\ce{Ca}}$ represent the density of H$_2$ molecules in vibrational state $\nu$, CaH molecules, and calcium atoms in the cell, respectively.

Buffer gas thermalization properties due to elastic collisions are included. The hot calcium atoms thermalize according to
\begin{equation}\label{eq:thermalization}
    \frac{dT_{\ce{Ca}}}{dt} = -\frac{R_{\ce{H2}}(T_{\ce{Ca,0}}-T_{\ce{H2}})}{\kappa_{\ce{H2}}} -\frac{R_{\ce{He}}(T_{\ce{Ca,0}}-T_{\ce{He}})}{\kappa_{\ce{He}}}
\end{equation}
which dictates the kinetic temperature of collisions. Here, $R_{\ce{BG}} = n_{\ce{BG}}\sigma^{el}_{\ce{BG}}\bar{v}_{\ce{BG}}$ is the elastic collision rate between calcium and buffer gas \ce{BG}, where $n_{\ce{BG}}$ is the density of the buffer gas, $\sigma^{el}_{\ce{BG}}$ is the elastic cross section, and $\bar{v}_{\ce{BG}}$ is the average relative velocity between calcium and the buffer gas. $T_{\ce{Ca}}$ is the temperature of the ablated calcium, $T_{\ce{BG}}$ is the buffer gas temperature, and $\kappa_{\ce{BG}} = (m_{\ce{Ca}} + m_{\ce{BG}})^2/(2m_{\ce{Ca}}m_{\ce{BG}})$. The elastic cross sections used in the model were experimentally extracted (see Supporting Information).

%\subsection{H$_2$ as a buffer gas coolant}
%\label{sec: buffer gas coolant}

\begin{figure}[ht!]
   \centering
   \includegraphics[scale=0.61]{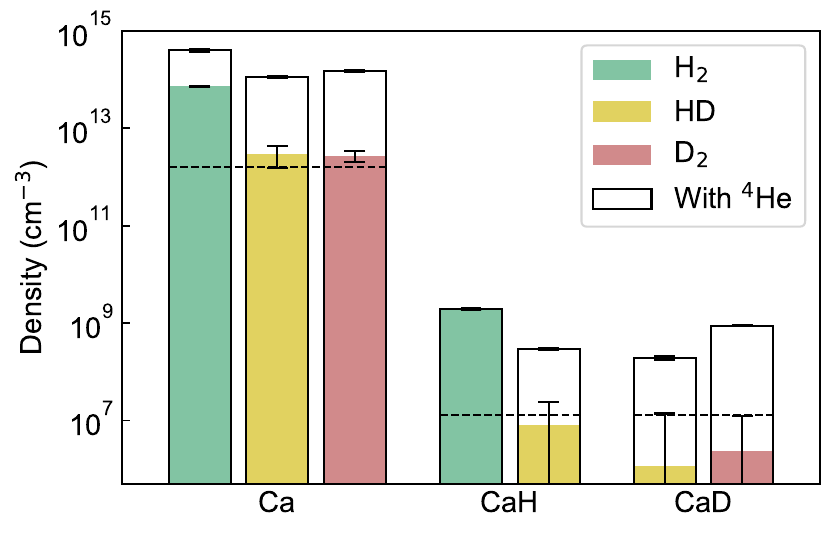}
   \caption{Time-averaged densities (0--1~ms after ablation) of Ca and CaH (or CaD) under different gas flow configurations. Reactant gases H$_2$, HD, or D$_2$ are supplied at $20$~SCCM, with He (if used) added at $8.8$~SCCM. Ablation energy is $19$~mJ per pulse. Green, yellow, and red bars represent results for H$_2$, HD, and D$_2$, respectively, while black-framed bars show the yields with He added. Dashed lines indicate the detection limits, defined as the lowest density detectable with $\text{signal-to-noise ratio (SNR)}>2$, below which measurements are statistically consistent with zero. Relative errors are larger when signals are close to the detection limit. Here the measured densities of CaH and CaD are those in the $X^2\Sigma^+(\nu=0,\ N=1,\ J=1/2,\ -)$ states.}
   \label{fig:H2_buffer_gas}
\end{figure}

A significant discrepancy is observed in the experimental yields of CaH or CaD depending on whether H$_2$, HD, or D$_2$ serves as the reactant gas. With H$_2$, substantial yields of thermalized calcium atoms and CaH molecules are detected, even in the absence of He as a buffer gas (Fig.~\ref{fig:H2_buffer_gas}, green). As a comparison, when HD or D$_2$ is supplied, the thermalized calcium yield drops significantly, and CaH (or CaD) production falls below our detection limits (Fig.~\ref{fig:H2_buffer_gas}, yellow and red). These findings highlight the advantages of H$_2$ as a reactant and cooling medium in CBGB systems.

These observations may be attributed to the exceptionally high saturated vapor pressure of H$_2$ at cryogenic temperatures. For example, at $6$~K the vapor pressure of H$_2$ reaches $\sim$$1.5$~mTorr, which is a few orders of magnitude above that of HD ($4.0\times10^{-2}$~mTorr) and D$_2$ (1.5$\times$10$^{-3}$~mTorr)~\cite{RN19,osti_4419030}. This high vapor pressure allows a sufficient density of H$_2$ in the gas phase, allowing more productive collisions with calcium atoms before adhering to the cell walls or exiting the cell, thereby facilitating efficient thermalization and chemical reactions. In comparison, HD and D$_2$ condense rapidly due to lower vapor pressures, resulting in insufficient collisions with calcium. Moreover, QCT simulations show that CaD formation in a D$_2$ environment is reduced by $30$\% relative to CaH production under H$_2$, which can be attributed to the larger mass of D$_2$. Further descriptions of the observed change in reactivity across isotopes may require a more rigorous quantum mechanical study.

Adding cryogenic He to the system increases the yields of thermalized calcium atoms and CaH molecules for all three reactants (black-framed bars in Fig.~\ref{fig:H2_buffer_gas}). This enhancement likely comes from the high efficiency of helium to thermalize the gas mixture and provide hydrodynamic entrainment~\cite{Hutzler_CR2012_BufferGasBeams}, thereby creating a more homogeneous environment in the cryogenic cell that improves detection of calcium and CaH. Therefore, introducing He into the system helps mitigate the constraints from the lower vapor pressures of HD and D$_2$. Note that the increase in CaH molecular yield is less than $22(1)$\%, substantially lower than the corresponding increase in thermalized calcium yield in the presence of He, which is typically greater than $440(40)$\%. This may result from a more rapid cooling of calcium atoms when cryogenic He is present, which subsequently lowers the efficiency of the chemical reaction.

%\subsection{Collisional cross sections}

Elastic collisions play a pivotal role in the thermalization process of CBGB systems.
Larger elastic cross sections correspond to higher collision rates, facilitating rapid thermal equilibration of translational degrees of freedom. Inert gases like He and Ne are commonly selected as buffer gases because of their relatively large elastic collisional cross sections with various molecular species, on the order of $10^{-14}$~cm$^2$. Motivated by the pronounced cooling effect observed with H$_2$, we investigate the elastic cross sections of H$_2$, HD, and D$_2$ in collisions with calcium and CaH.

\begin{figure}[t]
   \centering
   \includegraphics[scale=0.66]{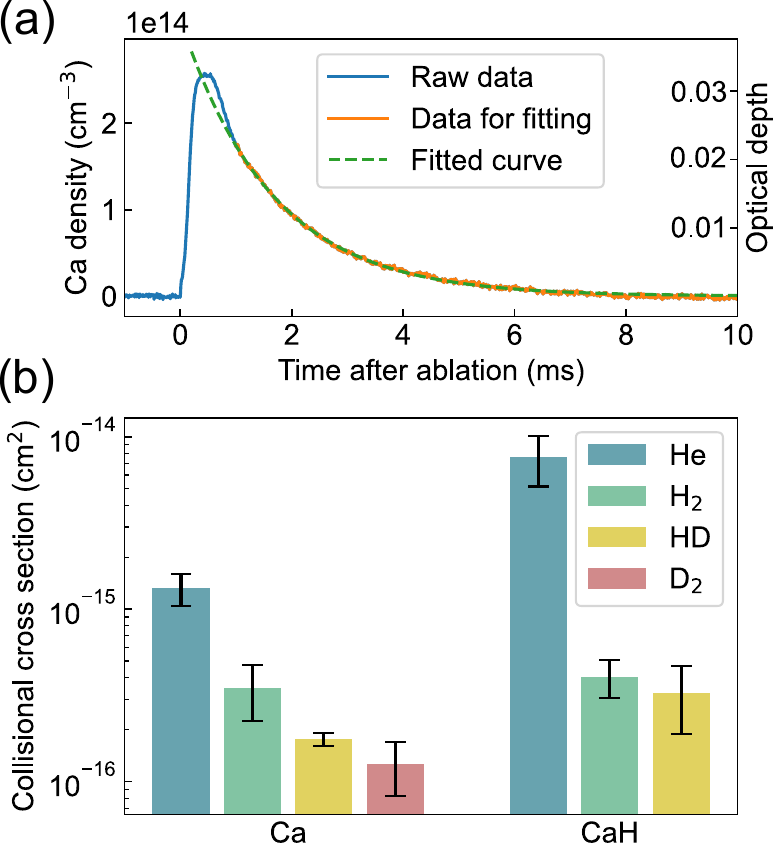}
   \caption{(a) A Ca density trace as a function of time after the ablation pulse (average of 30 shots). The blue curve is the original data, the orange line highlights the data used for fitting, and the green dashed line indicates the fitted single exponential decay result. This data is obtained with $8.8$~SCCM He flow. (b) Measured collisional cross sections of He, H$_2$, HD, and D$_2$ with Ca and CaH.}
   \label{fig:cross section}
\end{figure}

We use an approach adapted from Ref.~\cite{Hutzler_CR2012_BufferGasBeams}, where the elastic collisional cross section is extracted by tracking the exponential decays of in-cell signals over time. Details can be found in the Supporting Information. The data is presented in Fig.~\ref{fig:cross section}. In order to reduce measurement errors, we select multiple ablation spot and flow rate combinations and statistically average the results. Given the low densities of calcium and CaH (or CaD) when using the HD and D$_2$ gases, we average these measurements more extensively (see Supporting Information). Noticeable shifts in the background signal were observed during measurements involving H$_2$, HD, and D$_2$, likely due to sublimation of ice deposits on the cell windows. This was accounted for by including a constant offset in the exponential fit. Ultimately, we have measured the elastic cross section of H$_2$ with CaH to be $4.1(1.0)\times10^{-16}$~cm$^2$. This is high enough for efficient translational cooling in a regular buffer gas cell. We have also measured the ro-vibrational temperature of CaH with a H$_2$ buffer gas (see Supporting Information) which provided evidence for efficient cooling with H$_2$.

%\section{Discussion}

We have observed and quantified the remarkable chemical and thermal properties of H$_2$ molecules as a reactant and buffer gas in CBGB sources. Our study of the chemical production of CaH molecules demonstrates several notable phenomena that broaden the scope of cryogenic chemistry studies. This work also demonstrates a reliable method for generating cold CaH molecules in their ro-vibrational ground states, which shows a $\sim$$5$-fold improvement over previous approaches~\cite{Vazquez-Carson_2022,McNally_BaHSisyphus_2020, Dai2024PRR_CaD} and holds significant promise for producing ultracold trapped clouds of CaH and ultimately H atoms for quantum chemistry experiments and high-precision spectroscopy~\cite{Sun_PRR2023_Predissociation}. 

Producing CaH via a chemical reaction in the buffer gas has several distinct features over previous methods, which used a solid CaH$_2$ precursor. These differences come in the form of total CaH yield, forward velocity distribution and spread, and stability over time. We present a comparison across several references in Table~\ref{tab:comparison}. By performing a controlled chemical reaction to form CaH, we find a significant increase in the total molecular yield. While our peak forward velocity is larger than that of Ref.~\cite{Hsin-I_2011_CaH_Buffer_Gas}, which operated at a colder temperature of $1.8$~K with a two-stage cell, we find a similar velocity distribution to Ref.~\cite{Vazquez-Carson_2022}, suggesting the reaction does not have an adverse effect on the final velocity distribution. Finally, we see a significant decrease in the in-cell absorption time, likely due to the smaller collisional cross section (with CaH) of hydrogen than helium. Colder and slower beams may be generated by using a colder cell.

\begin{table}[t]
\begin{threeparttable}
\caption{Comparison of CaH molecular beam production from buffer gas cells. We compare the total CaH yield per pulse, the peak forward velocity, and the in-cell absorption time trace peak. Spreads are given in square brackets.}
\label{tab:comparison}
\centering
\setlength{\tabcolsep}{1pt}
% \begin{tabular}{l c c c}
\begin{tabular*}{\linewidth}{@{\extracolsep{\fill}}lccc}
\hline\hline
Reference & Yield & Peak velocity (m$/$s) & Absorption peak (ms) \\
\hline
Ref.~\cite{Hsin-I_2011_CaH_Buffer_Gas} & $5\times10^{8}$--$10^{9}$ & $64~[45]$ & $1~[3]$ \\ 
Ref.~\cite{weinstein2002magnetic} & $10^{10}$ & -- & -- \\
Ref.~\cite{Vazquez-Carson_2022} & $2 \times 10^{10}$ & $250~[100]$ & $1.5~[2]$ \\
This work & $10^{11}$  & $250~[100]$ & $0.3~[1]$\\ 
\hline\hline
\end{tabular*}
\end{threeparttable}
\end{table}

The successful production of CaH molecules despite the significant reaction barrier ($\sim$$2.7$~eV) demonstrates the effectiveness of H$_2$ as both a reactant and a cooling agent. This deep endothermic barrier can be overcome through the reaction network in which high-temperature Ca atoms facilitate CaH formation by transferring energy to vibrationally excite H$_2$. This observation challenges the conventional thermodynamic assumptions associated with endothermic reactions in CBGB setups. We also experimentally investigated the viability of enhancing the reaction with Ca excitation. We excited Ca to two possible states, $4s4p$ $^3P_1$ and $4s12s$ $^1S_0$, with the latter exceeding the reaction barrier of $\sim$2.7~eV. No enhancement was observed with the readily available laser powers ($\sim$$50$~mW). This implies the critical role of the reaction network as compared to the contribution of highly excited Ca atoms at the early stage ($\sim$$\mu$s) after laser ablation. The ability of H$_2$ to simultaneously facilitate chemical reactions and act as a buffer gas coolant highlights a dual functionality that has not been previously explored in depth.

We have simulated the reaction dynamics in a CBGB source using a reaction network model supplied with reaction rate constants from QCT simulations. This reaction model includes dependence of the reaction rate constants on the vibrational state of H$_2$ and successfully captures several qualitative trends observed in our experiments, such as the influence of translational temperature on reaction dynamics and the peak reaction efficiency of CaH. Due to the strong vibrational-translational coupling of this reaction, the formation of CaH is facilitated through vibrational excitations of H$_2$, subsequently opening reactive channels with other hot calcium atoms. One possibility is dissociation of CaH via collisions with calcium, for which the CaH density follows $\frac{d n_{\ce{CaH}}}{dt} = -k^dn_{\ce{Ca}}n_{\ce{CaH}}$ where $k^d$ represents the dissociation rate. While this expression depends on the calcium density, the low temperatures within the buffer gas cell and the relatively large binding energy of CaH suggests that dissociation does not play a major role. We also consider the possibility of three-body recombination reactions, such as the formation of the van der Waals molecule \ce{CaH-He} via \ce{CaH + He + He -> CaHHe + He}, since it depends on the CaH density. Using a classical threshold law for three-body recombination reactions, we estimate a recombination rate constant of $2.3\times10^{-32}$~cm$^6$$/$s for \ce{CaH-He} recombination at $6$~K~\cite{mirahmadi_classical_2021}, far too low to account for a significant drop in the ratio of densities. Additionally, the loss rate does not indicate any dependence on the buffer gas density (see reaction efficiencies at high densities in Fig.~\ref{fig:reaction}), whereas the diffusion coefficient is inversely proportional to the buffer gas density~\cite{Skoff_Tarbutt_2011_buffergas_temperature}. This suggests that losses due to diffusion do not play a major role within the parameters of the experiment. Instead, we find that losses due to Ca$^+$, which have an increased production at large ablation energies, reacting with CaH can effectively describe the discrepancy at large ablation energies or high densities (see Supporting Information).

The isotopic comparison reveals that H$_2$ outperforms D$_2$ and HD in both chemical reactivity and buffer gas cooling efficiency. This disparity can be attributed to the higher saturated vapor pressure of H$_2$ at cryogenic temperatures, as well as larger collisional cross sections. These properties ensure a more robust supply of cold H$_2$ molecules in the gas phase. This intriguing effect is partially captured by our reaction network model, mainly due to differences in mass across isotopes. Our observations indicate that it is necessary to account for the inherent quantum nature of atom-molecule collisions to quantitatively capture the isotopic effects on the reaction dynamics in a buffer gas cell.

This study advances the understanding of cryogenic chemical reactions and highlights the potential of H$_2$ as a versatile tool in CBGB systems. By combining experimental observations with computational modeling, we provide a comprehensive picture of the unique properties of H$_2$ and its isotopologues, paving the way for future explorations in this field.

This work was supported by the AFOSR MURI Grant No. FA9550-21-1-0069 and ONR Grant No. N00014-21-1-2644, and we acknowledge generous support by the Brown Science Foundation. R. K. and J. P.-R. acknowledge support by the AFOSR Grant No. FA9550-23-1-0202.

% \bibliography{references}


\end{document}

\title{Supporting Information for ``Chemistry in a Cryogenic Buffer Gas Cell''}

\author{Qi Sun}
\altaffiliation{These authors contributed equally to this work}
\affiliation{Department of Physics, Columbia University, New York, New York 10027, USA}

\author{Jinyu Dai}
\altaffiliation{These authors contributed equally to this work}
\altaffiliation{\\{\hypersetup{urlcolor=blue}\href{mailto:jd3706@columbia.edu}{jd3706@columbia.edu}}}
\affiliation{Department of Physics, Columbia University, New York, New York 10027, USA}

\author{Rian Koots}
\altaffiliation{These authors contributed equally to this work}
\affiliation{Department of Physics and Astronomy, Stony Brook University, Stony Brook, New York, New York 11794, USA}

\author{Benjamin C. Riley}
\affiliation{Department of Physics, Columbia University, New York, New York 10027, USA}

\author{Jes\'us P\'erez-R\'ios}
\email{jesus.perezrios@stonybrook.edu}
\affiliation{Department of Physics and Astronomy, Stony Brook University, Stony Brook, New York, New York 11794, USA}

\author{Debayan Mitra}
\email{debayanm@iu.edu}
\affiliation{Department of Physics, Columbia University, New York, New York 10027, USA}
\affiliation{Department of Physics, Indiana University, Bloomington, Indiana 47405, USA}

\author{Tanya Zelevinsky}
\email{tanya.zelevinsky@columbia.edu}
\affiliation{Department of Physics, Columbia University, New York, New York 10027, USA}

\date{\today}

\maketitle

\section{S1. Rotational temperature measurement}
\label{sec:density}

\begin{figure}[ht!]
   \centering
   \includegraphics[scale=0.63]{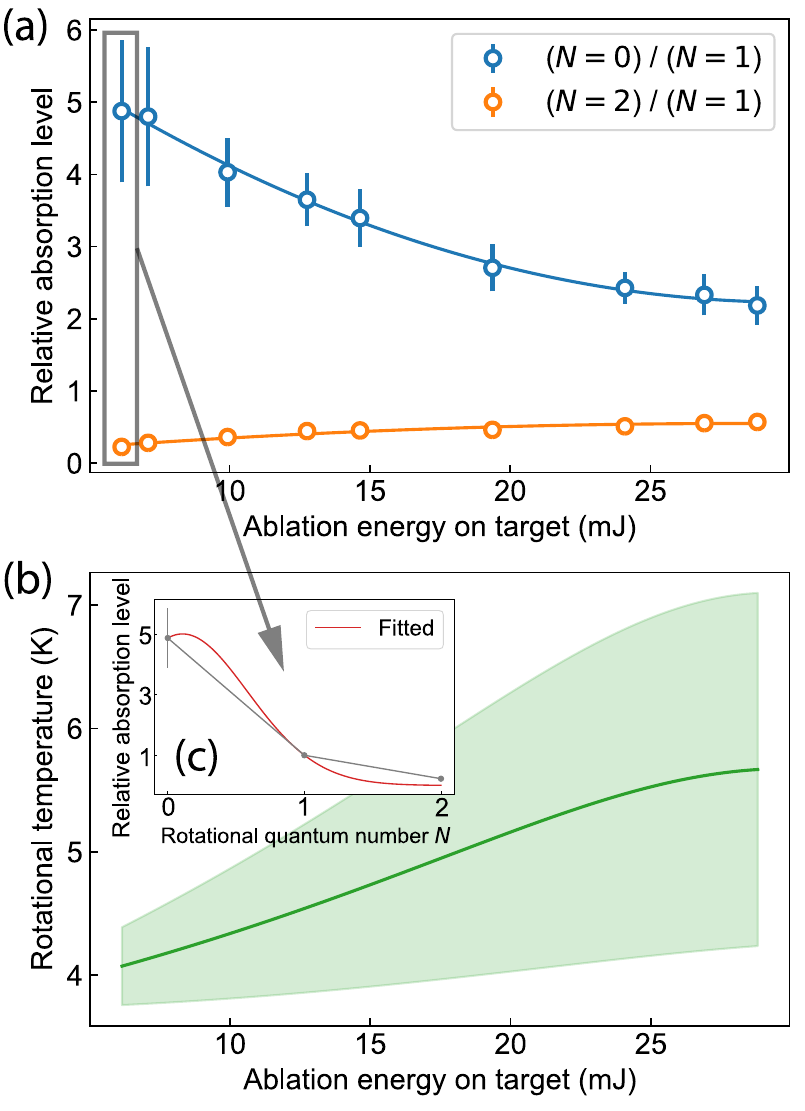}
   \caption{(a) Relative in-cell absorption for $X^2\Sigma^+(N=0)$ and $X^2\Sigma^+(N=2)$ states with respect to $X^2\Sigma^+(N=1)$, versus ablation pulse energy. Absorption for $X^2\Sigma^+(N=3)$ and higher ro-vibronic states is below the detection limit and therefore neglected. Solid lines are fitted quadratic functions to guide the eye. (b) Fitted rotational temperature versus ablation pulse energy. Temperatures are extracted from a least-squares fit according to Eq.~(\ref{eq:rotational temperature}), as shown in inset (c). The wide green band represents the 2-$\sigma$ uncertainties from the fit.}
   \label{fig:density_conversion}
\end{figure}

We measure the rotational temperature of the molecules by comparing the relative populations in different rotational states. Figure~\hyperref[fig:density_conversion]{\ref*{fig:density_conversion}(a)} shows the measured relative absorption of $X^2\Sigma^+(N=0)$ and $X^2\Sigma^+(N=2)$ states with respect to $X^2\Sigma^+(N=1)$ for varying ablation pulse energies. The transitions involved in these measurements are $A^2\Pi_{1/2}(\nu'=0,\ J'=1/2,\ -)\leftarrow X^2\Sigma^+(\nu''=0,\ N''=0,\ J''=1/2,\ +)$, $A^2\Pi_{1/2}(\nu'=0,\ J'=3/2,\ +)\leftarrow X^2\Sigma^+(\nu''=0,\ N''=1,\ J''=3/2,\ -)$, and $A^2\Pi_{1/2}(\nu'=0,\ J'=5/2,\ -)\leftarrow X^2\Sigma^+(\nu''=0,\ N''=2,\ J''=5/2,\ +)$, for the states with $N=0$, 1, and 2, respectively. Note that we use different transitions in the rest of the paper [$A^2\Pi_{1/2}(\nu'=0,\ J'=1/2,\ +)\leftarrow X^2\Sigma^+(\nu''=0,\ N''=1,\ J''=1/2,\ -)$], and numbers in Fig.~\hyperref[fig:density_conversion]{\ref*{fig:density_conversion}(a)} should not be directly interpreted as population ratios. We fit the relative absorption values to extract the corresponding rotational temperature via the expression
\begin{equation}
\label{eq:rotational temperature}
P\propto S\times d\times e^{-BN(N+1)/k_BT},
\end{equation}
where $P$, $S$, $d$, and $B$ denote the absorption, the H\"onl-London factor, the degeneracy of the respective hyperfine magnetic sublevels, and the ground-state rotational constant, respectively, and $k_B$ is the Boltzmann constant. The fitted rotational temperature as a function of ablation energy is shown in Fig.~\hyperref[fig:density_conversion]{\ref*{fig:density_conversion}(b)}, with the inset \hyperref[fig:density_conversion]{(c)} showing a sample fit. With the knowledge of the rotational temperature, we can use the population in the $X^2\Sigma^+(N=1)$ state to predict the population in all occupied rotational states. We find that the rotational states higher than $X^2\Sigma^+(N=2)$ can be neglected here since their population is very small at $\sim$$6$~K.

\section{S2. Ablation plume temperature measurement}
\label{sec:ablation_temperature}

\begin{figure}[ht!]
   \centering
   \includegraphics[scale=0.66]{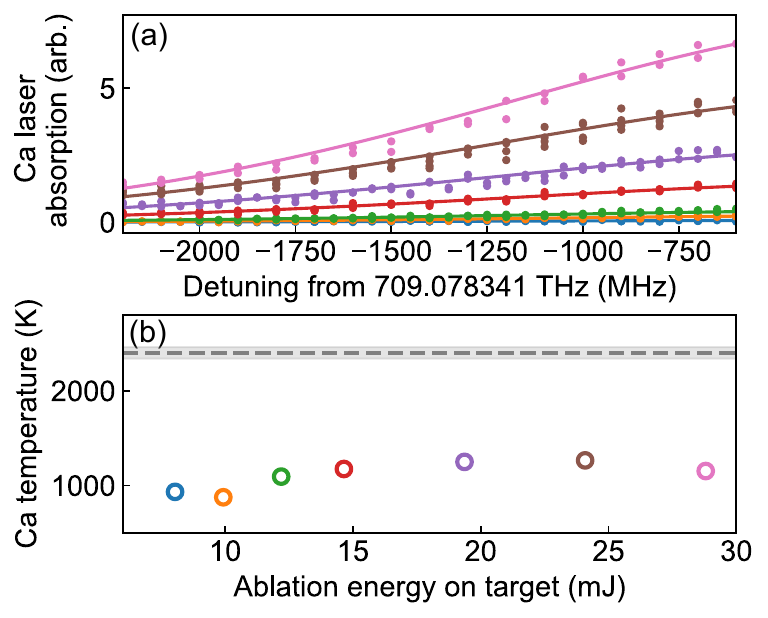}
   \caption{(a) Ca absorption spectra at different ablation energies. The data is fitted with the Gaussian function. (b) The Ca plume temperature versus ablation energy (error bars not visible at this scale). The gray shaded area represents the theoretical temperature {$2400(60)$~K} that optimally matches the experimental results in Fig.~2 of the main text. }
   \label{fig:plume_temperature}
\end{figure}

We have measured the dependence of the ablation plume temperature on the ablation energy when there is no gas flow in the system, as shown in Fig.~\ref{fig:plume_temperature}. The temperature of the ablation plume is determined by analyzing the measured absorption linewidth of the Ca $4s4p$ $^1P_1 \leftarrow4s^2$ $^1S_0$ transition at $423$~nm. While the Ca yield remains stable during the measurement, we sweep the frequency of the $423$~nm laser through the cell and observe a Gaussian-shaped absorption spectrum. Its linewidth is influenced by various line-broadening mechanisms which we carefully consider to identify the dominant contributor.

The natural linewidth of the transition is $\sim$$30$~MHz. Collisional broadening is negligible here due to the low density of the buffer gas in the cell, and power broadening is minimal due to the low intensity of the probe laser light. Consequently, the dominant broadening mechanism is attributed to the Doppler effect. The measured linewidth of $\sim$$1$~GHz is significantly larger than any of the other broadening contributions, allowing us to directly translate the linewidth to temperature.

It is important to note that the measured temperature represents a lower bound on the actual translational temperature of the ablation plume. In non-equilibrium systems such as the CBGB source, the translational temperature is anisotropic: the temperature along the direction of ablation is typically higher than the temperature perpendicular to it~\cite{Skoff_Tarbutt_2011_buffergas_temperature}. Based on our QCT simulations, a temperature of $2400(60)$~K provides the best fits to the experimental data. We therefore estimate this value to be near the average plume temperature in our system. Future experiments with more optical access could enable measurements along multiple spatial directions to provide a more precise characterization of the plume's anisotropic temperature distribution.

\section{S3. Extraction of collisional cross sections}
\label{sec:cross section}

Here we describe the determination of the elastic collisional cross sections. These measurements are based on the method in Ref.~\cite{Hutzler_CR2012_BufferGasBeams}. We assume that particles either stick to the cell walls or exit through the cell aperture via a diffusion process when the density of buffer gas is low. By monitoring the decay of the signal over time, we estimate the diffusion time constant and consequently derive the elastic collisional cross section.

The diffusion time constant is extracted by fitting an exponentially decaying function to the later part of the absorption time trace as illustrated in Fig.~4(a) of the main text. The elastic collisional cross section is then
\begin{equation}
\sigma = \frac{9\pi v_{\text{BG}} \tau}{16A_{\text{cell}}n_{\text{BG}}},
\end{equation}
where $v_{\text{BG}} = \sqrt{8 k_B T_0/\pi m_{\text{BG}}}$ is the expected speed of the buffer gas which follows the standard Maxwell-Boltzmann distribution, $\tau$ is the fitted diffusion time constant, $A_{\text{cell}}$ is the cross-sectional cell area of $\sim$$5$~cm$^2$, and $n_{\text{BG}}$ is the density of the buffer gas. The density is set by controlling the flow as $n_{\text{BG}} = 4f_{\text{BG}}/(v_{\text{BG}}A_{\text{aperture}})$, where $f_{\text{BG}}$ is the flow rate of the buffer gas into the system and $A_{\text{aperture}}$ is the aperture size of the buffer gas cell of $\sim$$0.25$~cm$^2$.

\begin{table}[t]
\centering
\begin{threeparttable}
\caption{Collisional cross sections (in units of $10^{-16}$~cm$^2$) for Ca and CaH with He, H$_2$, HD, and D$_2$.}
\label{tab:cross_sections}
\setlength{\tabcolsep}{4pt} % tighten for two-column layout
% \begin{tabular}{l c c c c}
\begin{tabular*}{\linewidth}{@{\extracolsep{\fill}}lcccc}
\hline\hline
 & He & H$_2$ & HD & D$_2$ \\
\hline
Ca   & $13.2(2.8)$ & $3.5(1.2)$ & $1.8(0.2)$ & $1.3(0.4)$ \\
CaH  & $77(25)$    & $4.1(1.0)$ & $3.3(1.4)$ & -- \\
\hline\hline
\end{tabular*}
\end{threeparttable}
\end{table}

To reduce systematic errors, we chose 3 different ablation spots and 2 different gas flow rates, and averaged the results to obtain the cross section values and the measurement error estimates. Overall, it takes 180 averages for He, 180 averages for H$_2$, 300 averages for HD, and 540 averages for D$_2$. The results are presented in Table~\ref{tab:cross_sections}.

\section{S4. QCT simulations}
\label{sec:QCT}

There have been numerous studies of the $\text{Ca}+\text{H}_2$ reaction which used pump-probe techniques, where Ca was introduced via a heat-pipe oven heated to around $900$~K and then excited to the $^1P$  and $^1D$ states to form CaH~\cite{Liu_Chen_Nien_Lin_1999, Song_Chen_Hsiao_Lin_Hung_2004, Chang_Chen_Hsiao_Chen_Lin_2005, Chen_Hsiao_Lin_2005}. On the theoretical front, six $^1$A$\prime$ potential energy surfaces (PES) of the $\text{Ca}+\text{H}_2$ system were presented~\cite{Kim_Lee_Lee_Jeung_2002}, and it was reported that these reactions are unlikely to occur from a single collision process due to large energy separations in the low energy region of the three PESs originating from Ca($4s3d$, $^{1}D$) and two PESs from Ca($4s4p$, $^{1}P$). In this work, we use the quasi-classical trajectory (QCT) method~\cite{truhlar1979,perezriosbook} via the PyQCAMS~\cite{pyqcams} software to calculate the energy-dependent cross sections of processes involved in collisions between ground state Ca and H$_2$, which are required to solve Eqs. (3)--(4) of the main text. The particle motion is treated classically, such that the trajectories are found using Hamilton's equations of motion. The internal states are treated according to the Bohr-Sommerfeld quantization rule, which quantizes the classical action into discrete energy levels. Thus, we gain insight into state-to-state cross sections of relevant processes, such as vibrational excitation or quenching of H$_2$, or a reaction leading to CaH formation, at variable collision energies. Our calculations agree with those of the most recent treatment of the $\text{Ca}(^1S)+\text{H}_2(\nu=0)$ reaction, which used a neural network method to construct a global PES and reported subsequent dynamics using a time-dependent wave packet method ~\cite{Liu_Chen_Zhao_Zhao_2024}. The energy-dependent cross section for a process $p$ is calculated via 
\begin{equation}\label{eq: sigma}
    \sigma_{p}(E_c) = 2\pi\int^{b_{max}}_{0} \mathcal{P}_{p}(E_c, b)bdb
\end{equation}
where $\mathcal{P}_p(E_c, b) = \frac{N_p}{N_t} \pm \frac{N_p}{N_t}\sqrt{\frac{N_t - N_p}{N_t}}$ represents the probability of process $p$ occurring from $N_t$ collisions for a given collision energy $E_c$ and impact parameter $b$. The uncertainty represents one standard deviation associated with the Boolean Monte Carlo process for calculating $\mathcal{P}_p(E_c, b)$. Here, process $p$ can represent rovibrational quenching, formation of a new molecule, or dissociation of the initial molecule. The integral is calculated after randomizing initial conditions for a sufficient number of trajectories. The reaction rate of the collision is then calculated as the product of the cross section and the collision velocity,
\begin{equation}\label{eq: rate}
    k_p(E_c) = \sigma_{p}(E_c)\sqrt{\frac{2E_c}{\mu}},
\end{equation}
where $\mu = m_{\ce{Ca}}m_{\ce{H2}}/(m_{\ce{Ca}} + m_{\ce{H2}})$ is the reduced mass of the colliding particles.
For an appropriate Maxwell-Boltzmann distribution of energies, we calculate the temperature-dependent rate constants of these processes,
\begin{equation}\label{eq: temprate}
    k_p(T) = \frac{2}{\left(k_B T\right)^{3/2}}\int k_p(E_c)\sqrt{\frac{E_c}{\pi}} e^{-E_c/k_BT}dE_c.
\end{equation}

\begin{table}[t]
\begin{threeparttable}
\caption{Parameters used in the Morse potential for pairwise interactions. The H$_2$ potential was calculated with the full configuration interaction method, and the CaH potential was obtained from Ref.~\cite{shayesteh2017} to which we fit the Morse parameters.}
\label{tab:params}
\centering
\setlength{\tabcolsep}{4pt}
% \begin{tabular}{lccc}
\begin{tabular*}{\linewidth}{@{\extracolsep{\fill}}lccc}
\hline\hline
 & $D_e$ ($E_H$) & $r_e$ ($a_0$) & $\alpha$ ($a_0^{-1}$) \\
\hline
H$_2$ ($X^1\Sigma_g^+$)   & $0.165$  & $1.40$ & $1.06$ \\
CaH ($X^2\Sigma^+$)      & $0.0653$ & $3.79$ & $0.691$ \\
\hline\hline
\end{tabular*}
\end{threeparttable}
\end{table}

The Hamiltonian governing the collision is given by
\begin{equation}
    H = \sum_{i=1}^{3} \frac{\vec{p_i}^2}{2m_i} + V\left(\vec{r_1},\vec{r_2}, \vec{r_3}\right).
\end{equation}
The three-body potential is approximated as a pairwise additive potential,
\begin{equation}\label{eq: potential}
    V\left(\vec{r_1},\vec{r_2}, \vec{r_3}\right) = V(r_{12}) + V(r_{23}) + V(r_{31}).
\end{equation}
To approximate two-body interactions, we use the Morse potential, given by $V (r) = D_{e}(1-e^{-\alpha(r-r_e)})^{2} - D_{e}$, where $r$ is the interparticle distance, $D_{e}$ is the dissociation energy, and $r_e$ is the equilibrium distance of the diatomic molecule. The potential parameters for H$_2$ and CaH are listed in Table~\ref{tab:params}.

\begin{figure}[ht!]
    \centering
    \includegraphics[scale=0.46]{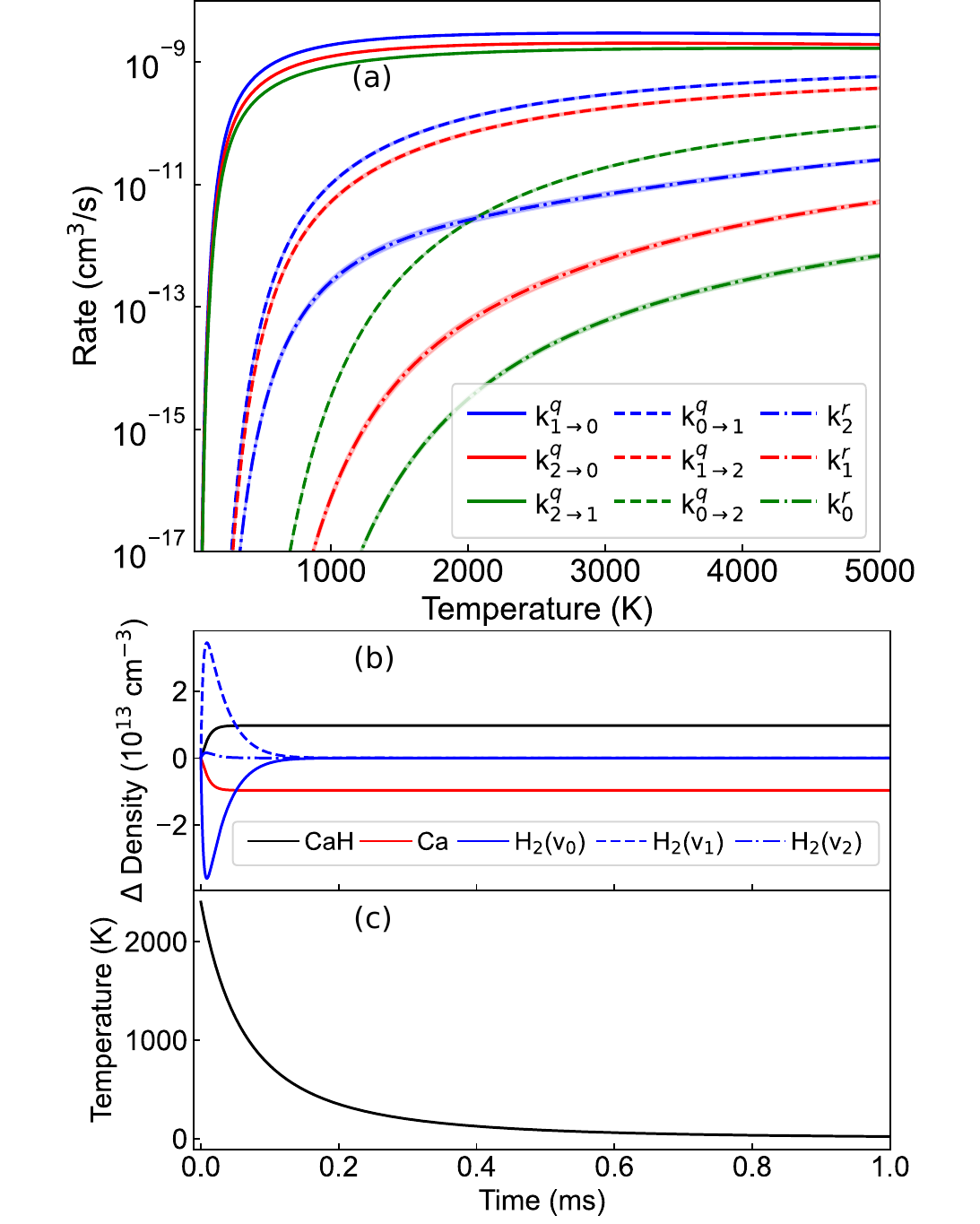}
    \caption{(a) QCT calculated temperature-dependent state-to-state vibrational rate constants for H$_2$ ($k^q_{\nu \rightarrow \nu^{\prime}}$), where the subscripts denote the initial and final vibrational states, and formation of CaH ($k^r_\nu$), where the subscript denotes the initial vibrational state of H$_2$. (b) Sample solution to the reaction model at $2400$~K showing the change in number density over time relative to the initial densities. The Ca and CaH densities are scaled up by a factor of $500$ for visualization. (c) Kinetic temperature of Ca over time, decaying due to elastic collisions with buffer gases H$_2$ and He. }
    \label{fig:rates}
\end{figure}

Figure~\hyperref[fig:rates]{\ref*{fig:rates}(a)} presents the temperature-dependent rate constants that were used in our reaction model. The choice of states included in the reaction model was made by mapping the rate constants and characteristic timescales of vibrational quenching of H$_2$ at a collision temperature of $2400$~K, where the density of H$_2$ is $\sim$$7.7\times10^{14}$~cm$^{-3}$, corresponding to a $20$~SCCM flow rate in the buffer gas cell. At this temperature and density, we find that throughout the reaction, $\sim$$3.7$\% of the total H$_2$ density is expected to be excited to the first vibrational state ($\nu=1$), and only $\sim$$0.02$\% to ($\nu=2$). Thus, the inclusion of these vibrationally excited states of H$_2$ are sufficient to describe the cell dynamics. While there is a trend of higher vibrational states leading to larger reaction rate constants at low temperatures, excitation of H$_2$ to vibrational states larger than ($\nu=2$) is highly unlikely under buffer gas cell conditions.

We do not include rotational excitations of H$_2$ in the model, as they show little influence on the relevant reactions. In Table~\ref{tab:rotation} we show the impact of the initial rovibrational state of H$_2$ on the calculated rate coefficients of CaH formation and state-to-state quenching of H$_2$ at a sample temperature of 3,000~K. We find that the calculated reaction rate constants are largely independent of the initial rotational state of the reactants but are dependent on the initial vibrational states. In addition, the efficient rotational quenching of H$_2$ in He buffer gas restricts the lifetime of excited rotational states. Therefore, we only consider the ground rotational state of H$_2$ throughout the reaction ($j_i=0$), but include the state-to-state vibrational inelastic rate constants by summing over all final rotational states. As H$_2$ becomes vibrationally excited, the endothermicity of the reaction $\text{Ca}+\text{H}_2$ is reduced by $\sim$$0.55$~eV, corresponding to the vibrational spacing of H$_2$, making CaH formation more probable at buffer gas temperatures. More importantly, vibrationally excited H$_2$ molecules exhibit larger bond lengths, which plays a significant role in reactive collisions~\cite{Kim_Lee_Lee_Jeung_2002}. In other words, the reaction is facilitated by vibrational excitation of H$_2$ since vibrational-translational coupling is shown to be strong for this system.

\begin{table}[t]
\begin{threeparttable}
\caption{Influence of the initial rovibrational state ($\nu$, $j$) of H$_2$ on the CaH formation rate constant ($k^r_{\nu}$), and on the state-to-state H$_2$ vibrational quenching rate constants, where the vibrational state is lowered by one ($k^q_{\nu~\rightarrow~\nu-1}$) or raised by one ($k^q_{\nu~\rightarrow~\nu+1}$), at a collision temperature of 3,000~K. We sum over all rotational states, and present the rate constants in units of cm$^3$$/$s.}
\label{tab:rotation}
\centering
\setlength{\tabcolsep}{4pt}
% \begin{tabular}{lccc}
\begin{tabular*}{\linewidth}{@{\extracolsep{\fill}}lccc}
\hline\hline
($\nu$, $j$)
& $k^r_{\nu}$($\times10^{-14}$)
& $k^q_{\nu\rightarrow \nu-1}$($\times10^{-9}$)
& $k^q_{\nu\rightarrow \nu+1}$($\times10^{-10}$) \\
\hline
(0, 0) & $7.59 \pm 0.69$ & --                 & $2.97 \pm 0.10$ \\
(0, 2) & $7.00 \pm 0.63$ & --                 & $2.63 \pm 0.10$ \\
(1, 0) & $59.3 \pm 7.11$ & $3.57 \pm 0.04$    & $1.78 \pm 0.08$ \\
(1, 2) & $53.7 \pm 7.42$ & $3.42 \pm 0.04$    & $1.63 \pm 0.08$ \\
\hline\hline
\end{tabular*}
\end{threeparttable}
\end{table}

The initial Ca density and collision temperature are dictated by the ablation laser power, and serve as inputs for the reaction model. After finding the steady-state solution to Eqs.~(3)-(4) of the main text, we report the expected density of CaH molecules forming in the buffer gas cell. A sample reaction process with the initial Ca density of $9.2\times10^{13}$~cm$^{-3}$ and H$_2$ density of $7.5\times10^{14}$~cm$^{-3}$ is shown in Fig.~\hyperref[fig:rates]{\ref*{fig:rates}(b)}, where the change in density of each species is plotted. Thermalization of Ca is shown in Fig.~\hyperref[fig:rates]{\ref*{fig:rates}(c)} for the same sample reaction. CaH formation is complete in $\lesssim$$0.2$~ms, implying that all the chemistry happens before detection and that the rapid thermalization of Ca subsequently restricts the reaction.

The main source of uncertainty in the model (shaded area around the solid blue and red lines in Fig.~2 of the main text representing one standard deviation) is from the QCT rate constants. To propagate the error through the model, we sample 100 rate constants from the associated error distribution and solve the steady-state reaction network. We then collect statistics on the density of each species at each calculation step.

\section{S5. Loss channel}\label{sec:losses}
\begin{figure}[t]
    \centering
    \includegraphics[scale=0.56]{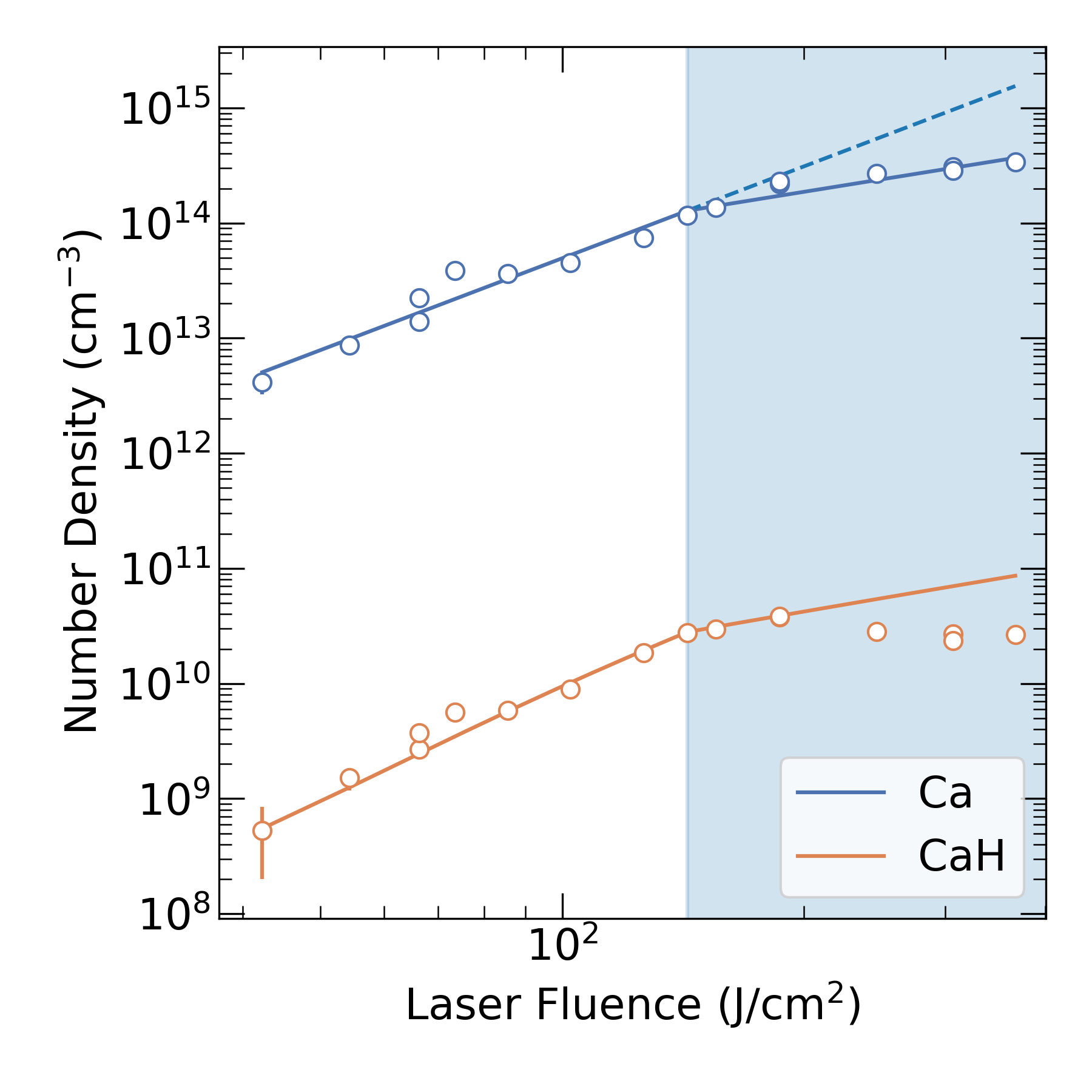}
    \caption{Number density of Ca and CaH as a function of ablation laser fluence, demonstrating a change in power law behavior. Ca density serving as the input of the model is represented by the solid blue curve, with experimentally measured Ca densities as blue open circles. The CaH density is represented by the orange curve, and is the output of the reaction model. The orange open circles represent experimentally measured CaH densities. The data points were taken with $40$~SCCM of H$_2$ and $2.2$~SCCM of He flowing through the cell. The shaded region is where plasma shielding is expected to occur. The dashed line represents the predicted Ca density without plasma shielding, and the difference between the dashed and solid lines indicates the amount of Ca$^+$ ions formed.}
    \label{fig:shielding}
\end{figure}

At high ablation energies, the theoretical prediction of CaH production deviates from experimental measurements. It is well known that ablation of materials suffers from plasma shielding at large ablation energies~\cite{Russo_2002}, when the plume becomes ionized. The freed electrons absorb the incoming light, which presents as a decreased ablation efficiency. This effect appears when plotting the number density of ablated atoms against the laser fluence in our system, as shown in Fig.~\ref{fig:shielding}. At a laser fluence of $\sim$$150$~J$/$cm$^2$, the ablation efficiency of Ca shows a clear change in trend. At this threshold, marked by the shaded region, the deviation between experiment and theory begins, as shown by the orange CaH density curve. This is evidence of a new reactive mechanism in the system which is linked to the plasma shielding effect.

\begin{figure}[ht!]
    \centering
    \includegraphics[scale=0.42]{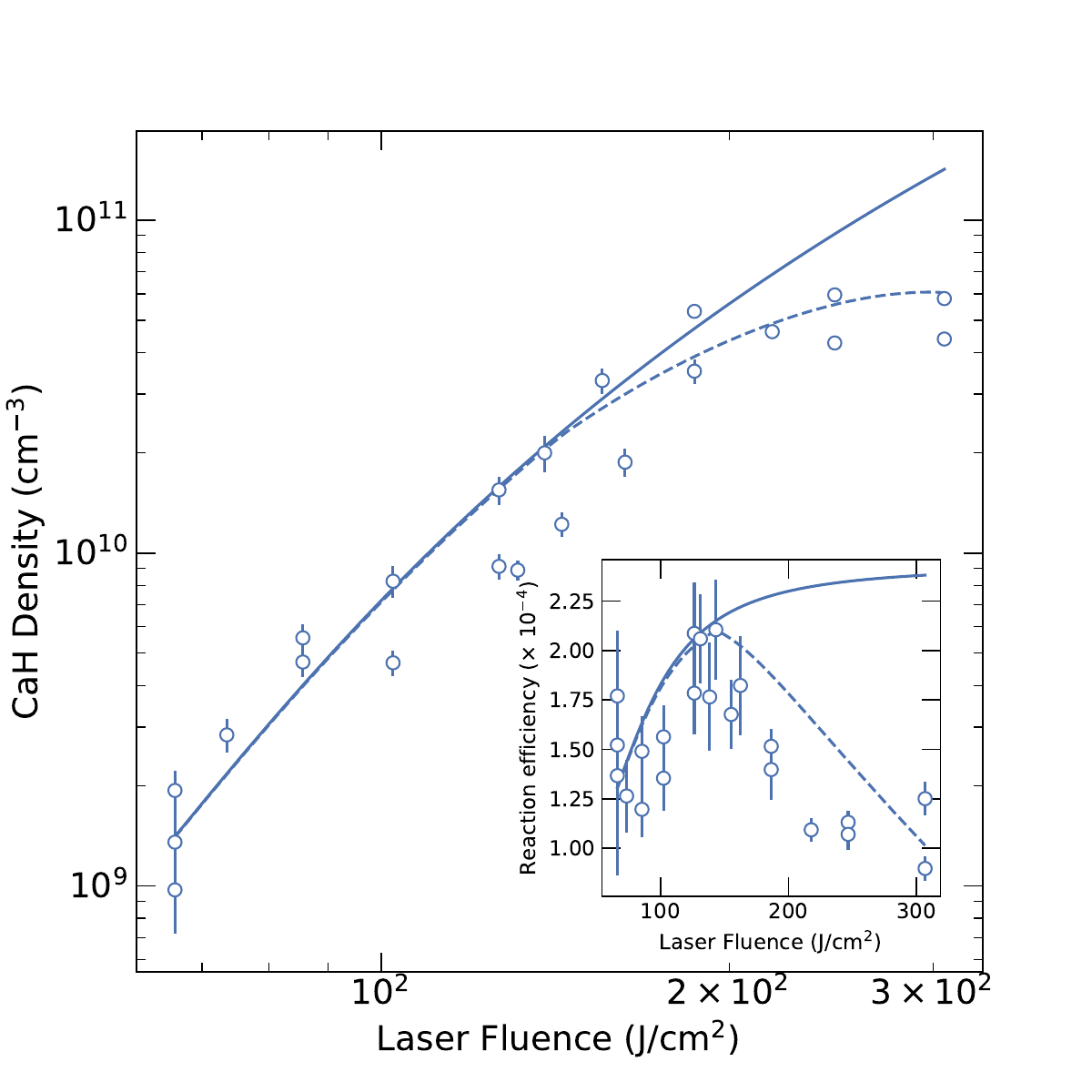}
    \caption{Effect of Ca$^+$ ion production due to plasma shielding on CaH number density as a function of ablation laser fluence. The data points are a combination of $10$~SCCM and $20$~SCCM H$_2$, with $2.2$~SCCM of He flow. The solid line is the reaction model output without Ca$^+$ present, and the dashed line is with Ca$^+$ produced by laser ablation. The inset shows the reaction efficiency, defined as the ratio of CaH and Ca densities. The errors of the theoretical predictions are not shown but resemble those in Fig.~2 of the main text.}
    \label{fig:ion_effect}
\end{figure}

At the onset of plasma sheilding, we expect an increased production of Ca$^+$ ions by the ablation laser, which opens possible loss channels:
\begin{align}
    \label{eq:ion_loss}
    \ce{Ca+ + CaH &-> Ca2+ + H},\\
    \ce{Ca+ + CaH &-> CaH+ + H}.
    \label{eq:ion_loss2}
\end{align}
The rate coefficient of these reactions can be estimated with a simple capture model which takes into account the polarizability and permanent dipole moment of CaH~\cite{Dugan_Magee_1967}. For polarizability $\alpha$ and dipole moment $\mu_D$, the energy-dependent rate coefficient for losses due to ionic collisions, $k_{\text{loss}}(E)$, is given by
\begin{equation}
    k_{\text{loss}}(E_c) = \pi\left(2\sqrt{\frac{\alpha}{\mu}} + \sqrt{\frac{2\mu_D^2}{\mu E_c}}\right)
\end{equation}
where $\mu$ is the reduced mass of the \ce{Ca+-CaH} system, and we have assumed there is sufficient time for the dipole to align with the electric field of the ion. Beyond the plasma shielding threshold, we allow the density of ions to grow proportionally to the difference between the two power laws describing Ca ablation shown in Fig.~\ref{fig:shielding} (dashed and solid lines). Since the number of ions produced by ablation is a free parameter, we find agreement if the fraction of Ca$^+$ ions produced by ablation is $\sim$1$/$7,000 of the difference between the dashed and solid lines in Fig.~\ref{fig:shielding}. In other words, the apparent reduced ablation efficiency of Ca is accompanied by a proportional increase in the production of Ca$^+$ ions.

In Fig.~\ref{fig:ion_effect} we show the CaH density before and after the inclusion of losses due to Ca$^+$ ions represented as Eqs.~(\ref{eq:ion_loss})--(\ref{eq:ion_loss2}), where we now find agreement at large ablation energies. These results indicate that the production of ions at large ablation energies has a significant loss effect on the formation of CaH in the buffer gas cell.

% \bibliography{references}
